# Fractional quantum ferroelectric control of spin-valley locking and valley Hall effects in altermagnetic monolayer $Cr_2S_2$

Tao Yao, Quan Shen, Jiansheng Dong*, Jianing Tan*

*Department of Physics, Jishou University, Jishou 416000, Hunan, China*

*Corresponding authors: jsdong@jsu.edu.cn, jianing_tan@163.com

Fractional quantum multiferroics, arising from the coupling between fractional quantum ferroelectricity (FQFE) and altermagnetism (AM), provide a promising platform for nonvolatile control of momentum dependent spin splitting in systems with zero net magnetization. However, extending this FQFE-AM coupling to valley degrees of freedom and Berry curvature driven valley Hall effects remains largely unexplored. Here, using first-principles calculations, we demonstrate that monolayer $Cr_2S_2$ realizes a two dimensional FQFE-AM platform with two switchable FQFE states connected by composite symmetry operations combining a fractional lattice translation with time reversal or parity-time reversal. We show that FQFE switching reverses the AM spin-polarized band structure and interchanges the spin characters of the X and Y valleys without rotating the Néel vector, thereby enabling polarization switchable spin-valley locking. Moreover, the two FQFE states exhibit reversed Berry curvature distributions, which, together with the switched spin-valley locking, enable polarization controlled valley Hall effects under both electron and hole doping. These results demonstrate a symmetry based mechanism for nonvolatile electrical control of AM spin splitting, spin-valley locking, and valley Hall effects, offering a general route toward low-power valleytronic devices based on FQFE-AM coupling.

*Introduction.* Fractional quantum ferroelectricity (FQFE) has recently emerged as a new route to switchable ferroelectric polarization beyond the conventional symmetry constraints of polar point groups, thereby expanding the materials space of ferroelectrics [1,2]. In contrast to conventional ferroelectricity driven by symmetry lowering polar lattice distortions, FQFE can arise from fractional lattice displacements of atoms between symmetry equivalent sites, thereby enabling switchable polarization in crystals that are nonpolar in the conventional sense [3,4]. This mechanism further provides a new degree of freedom for coupling electric polarization with magnetism. In this context, altermagnetism (AM) serves as an ideal magnetic counterpart for constructing fractional quantum multiferroics (FQMFs), because both FQFE and AM are closely tied to lattice symmetry, while the FQFE switching can further involve a fractional lattice translation $\tau$ [5]. AMs combine key advantages of antiferromagnets, such as vanishing net magnetization, absence of stray fields, and high magnetic stability, with symmetry protected momentum dependent spin splitting [6-10]. Crucially, because this spin splitting is intrinsically tied to the crystal symmetry and magnetic sublattice arrangement, it can be manipulated by structural or polarization switching without requiring rotation of the Néel vector. When FQFE is coupled to AM, the resulting FQMF state exhibits a symmetry enforced magnetoelectric coupling [5]. Symmetry analysis reveals that this coupling is governed by the composite operations $\mathcal{T}\tau$ or $\mathcal{PT}\tau$, where $\mathcal{T}$ and $\mathcal{P}$ denote time reversal and $\mathcal{P}$ spatial inversion operations, respectively, and $\mathcal{PT}$ denotes the combined parity-time operation. Under these symmetry constraints, FQFE polarization switching can reverse the AM spin splitting without reorienting the Néel vector, offering a symmetry based route toward electrically controlled spintronic and valleytronic functionalities.

The valley degree of freedom, associated with energetically degenerate but momentum-separated band extrema, has emerged as a promising information carrier for encoding, processing, and transport [11-14]. In conventional hexagonal materials, valley physics is usually associated with inequivalent valleys located at the Brillouin-zone corners, which are related by $\mathcal{T}$ symmetry in nonmagnetic systems [15,16]. By contrast, in square lattice two dimensional AM systems, the relevant crystal valleys can appear at symmetry related high symmetry momenta, with their degeneracy protected by crystalline symmetries rather than by conventional $\mathcal{T}$ related valley pairing [17-19]. When these valleys carry opposite spin characters, spin-valley locking emerges, enabling the coupled manipulation of spin and valley information. More importantly, if these symmetry related valleys exhibit opposite Berry curvatures, an in-plane electric field can drive valley dependent anomalous transverse velocities, giving rise to the valley Hall effect [20]. These considerations raise a central question for FQMFs: can FQFE polarization switching go beyond reversing AM spin splitting and simultaneously switch the spin-valley locking pattern and Berry curvature driven valley Hall response?

In this Letter, we demonstrate FQFE polarization controlled spin-valley locking and valley Hall responses in monolayer $Cr_2S_2$, a two dimensional FQFE-AM platform. We identify two switchable FQFE states, denoted as $h_1$ and $h_2$, which are connected by the composite symmetry operations $\mathcal{T}\tau$ or $\mathcal{PT}\tau$. This symmetry relation allows FQFE polarization switching to reverse the AM spin-polarized band structure without reorienting the Néel vector. As a consequence, the spin characters of the X and Y valleys are interchanged between the two FQFE states, leading to polarization switchable spin-valley locking. Furthermore, $h_1$ and $h_2$ exhibit reversed Berry curvature distributions, enabling nonvolatile control of the valley Hall effects under both electron and hole doping. Our results extend the functionality of FQMFs from electrically switchable AM spin splitting to FQFE polarization controlled spin-valley locking and Berry curvature driven valley Hall effects.

*Computational details.* First-principles calculations were performed within density functional theory (DFT) using the Vienna *ab initio* simulation package (VASP) [21,22]. The projector augmented-wave (PAW) method was employed to describe the electron-ion interactions, and the exchange-correlation potential was treated using the Perdew-Burke-Ernzerhof (PBE) functional within the generalized gradient approximation (GGA) [23]. A vacuum layer of 20 Å was introduced along the out-of-plane direction to avoid spurious interactions between periodic images. Following previous studies, the localized Cr 3*d* electrons were treated using the GGA+U method, with an effective Hubbard parameter of $U_{\mathrm{eff}}$ = 2.26 eV [24-26]. The two dimensional Brillouin zone was sampled using a 14×14×1 Monkhorst-Pack *k*-point mesh. The plane-wave cutoff energy and total energy convergence criterion were set to 500 eV and $10^{-6}$ eV, respectively. All structures were fully relaxed until the residual forces on each atom was less than 0.01 eV/Å. The electric polarization was calculated using the Berry-phase method [27-30]. The dynamical stability of monolayer $Cr_2S_2$ was evaluated by calculating the phonon dispersion using the PHONOPY code with the finite-displacement method on a 4×4×1 supercell [31]. Thermal stability was further examined through *ab initio* molecular dynamics (AIMD) simulations performed on the same supercell [32]. The climbing-image nudged elastic band (CI-NEB) method was employed to determine the minimum-energy switching paths and energy barriers [33]. The Berry curvature was calculated using the VASPBERRY code [34,35].

*Results and discussion.* Within the modern theory of polarization, an adiabatic lattice displacement changes the Berry-phase polarization, which is defined modulo a polarization quantum [27,36]. In FQFE, such displacements correspond to fractional lattice translations between symmetry equivalent configurations, allowing switchable electric polarization even in systems beyond conventional polar point groups. When AM order is coupled to an FQFE lattice, the resulting FQMF state provides a natural platform for coupling electric polarization with momentum dependent spin splitting [5]. Unlike conventional magnetoelectric mechanisms that often rely on spin reorientation, the spin splitting in AM is symmetry locked to the crystal structure and magnetic sublattice arrangement. FQFE switching can therefore reverse the spin-polarized band structure while preserving the Néel-vector orientation. If the relevant band extrema form symmetry-related crystal valleys, this reversal also switches the valley spin characters, producing a polarization controllable spin-valley locking pattern.

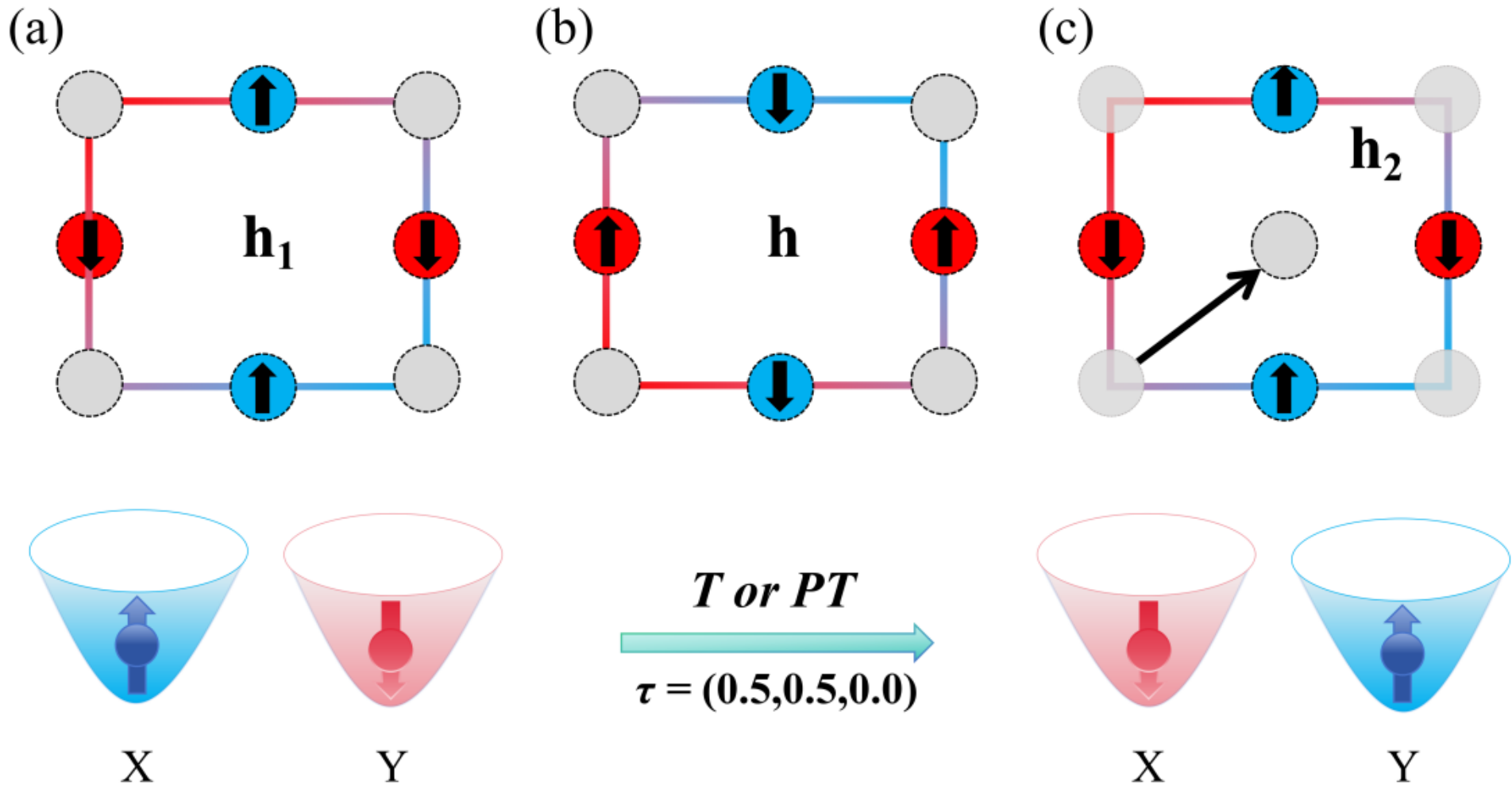

FIG.1 Schematic illustration of FQFE controlled spin-valley locking in an FQMF. (a) FQFE state $h_1$ with collinear AM order. Gray spheres denote nonmagnetic atoms, while blue and red spheres denote magnetic atoms with opposite spin orientations. The bottom schematic diagrams indicate opposite spin characters at the X and Y valleys. (b) Auxiliary $h$ state obtained from $h_1$ state by applying $\mathcal{T}$ or $\mathcal{PT}$ reversal. (c) FQFE state $h_2$ generated from $h$ state by a fractional lattice translation $\tau$ = (0.5, 0.5,0.0), with black arrows indicating the associated atomic displacements. The composite operation $\mathcal{T}\tau$ or $\mathcal{PT}\tau$ reverses the spin-polarized band structure and interchanges the X/Y valley spin characters without rotating the Néel vector.

We first illustrate this mechanism using the phenomenological two dimensional FQMF model shown in Fig. 1. Figure 1(a) depicts the FQFE state $h_1$, in which the gray atoms are nonmagnetic, whereas the blue and red atoms carry opposite spin orientations and form a collinear AM configuration. The displacement of the nonmagnetic atoms from their high symmetry lattice positions gives rise to FQFE polarization. Meanwhile, owing to the momentum dependent spin splitting of AM, the electronic states at the X and Y valleys carry opposite spin characters. As indicated by the valley schematics below the atomic structure, the $h_1$ state therefore exhibits spin-valley locking, with opposite spin components localized at the two valleys.

The symmetry relation between the two FQFE states can be understood from Figs. 1(a)-1(c). Starting from $h_1$ state, applying $\mathcal{T}$ or $\mathcal{PT}$ reversal produces the auxiliary configuration $h$ [Fig. 1(b)], which does not correspond to a stable intermediate state along the actual switching path, but rather serves to reveal the symmetry connection between the two FQFE states. A subsequent fractional lattice translation $\tau$=(0.5, 0.5, 0.0) transforms h into the second state $h_2$ [Fig. 1(c)]. Thus, the two physical FQFE states are connected by a composite symmetry operation: $h_2 = T\tau h_1$ or $h_2 = PT\tau h_1$. The black arrow in Fig. 1(c) indicates the atomic displacement associated with this fractional translation, which switches the FQFE polarization between $h_1$ and $h_2$ states.

Within the framework of spin-group symmetry, the above symmetry relations impose direct constraints on the spin-resolved band structures. For the $\mathcal{T}\tau$ related states, the band energies satisfy [6,7,37]

$$\epsilon_n^2(\mathbf{k},\mathbf{s}) = [C_2 \| \mathcal{T} | \tau]\epsilon_n^1(\mathbf{k},\mathbf{s}) = \epsilon_n^1(-\mathbf{k},-\mathbf{s}) \xlongequal{[\overline{C_2} \| \mathcal{T}]} \epsilon_n^1(\mathbf{k},-\mathbf{s}) \tag{1}$$

whereas for the $\mathcal{PT}\tau$ related states, one obtains

$$\epsilon_n^2(\mathbf{k},\mathbf{s}) = [C_2 \| \mathcal{PT} | \tau]\epsilon_n^1(\mathbf{k},\mathbf{s}) = \epsilon_n^1(\mathbf{k},-\mathbf{s}) \tag{2}$$

where $\epsilon_n^1(\mathbf{k},\mathbf{s})$ and $\epsilon_n^2(\mathbf{k},\mathbf{s})$ denote the band structures of $h_1$ and $h_2$, respectively. The $\overline{C_2} \| \mathcal{T}$ symmetry originates from the spin-only group. These symmetry relations show that FQFE polarization switching reverses the spin-polarized band structure while leaving the Néel vector unchanged.

This reversal directly interchanges the spin characters of the X and Y valleys, thereby reversing the spin-valley locking pattern. In the $h_1$ state, the X and Y valleys carry opposite spin characters, forming a specific spin-valley locking pattern. After FQFE switching to $h_2$ state, this pattern is reversed: the spin-up X valley and spin-down Y valley in $h_1$ state become the spin-down X valley and spin-up Y valley in $h_2$ state, as illustrated in Figs. 1(a) and 1(c). Therefore, the model establishes a symmetry-based mechanism by which FQFE switching simultaneously reverses AM spin splitting and switches the X/Y valley spin characters. This result extends the FQFE-AM coupling from electrically switchable spin splitting to polarization switchable spin-valley locking, providing the physical basis for polarization controlled valley dependent responses in FQMFs.

Having established the symmetry mechanism for FQFE controlled spin-valley locking, we next examine monolayer $Cr_2S_2$ as a realistic two dimensional FQFE-AM platform. Monolayer $Cr_2S_2$ crystallizes in the space group (*P4/mmm*) (No. 123) and adopts a tetragonal lattice with an S-Cr-S trilayer configuration, as shown in the inset of Fig. 2(a), where the blue and green spheres denote Cr and S atoms, respectively. We first assessed its thermal stability using AIMD simulations at 300 K. As shown in Fig. 2(a), the total energy exhibits only small fluctuations during the simulation, and the crystal structure remains intact, indicating good thermal stability. Dynamic stability was further confirmed by the phonon dispersion spectrum in Fig. 2(d), where no imaginary frequencies appear throughout the Brillouin zone.

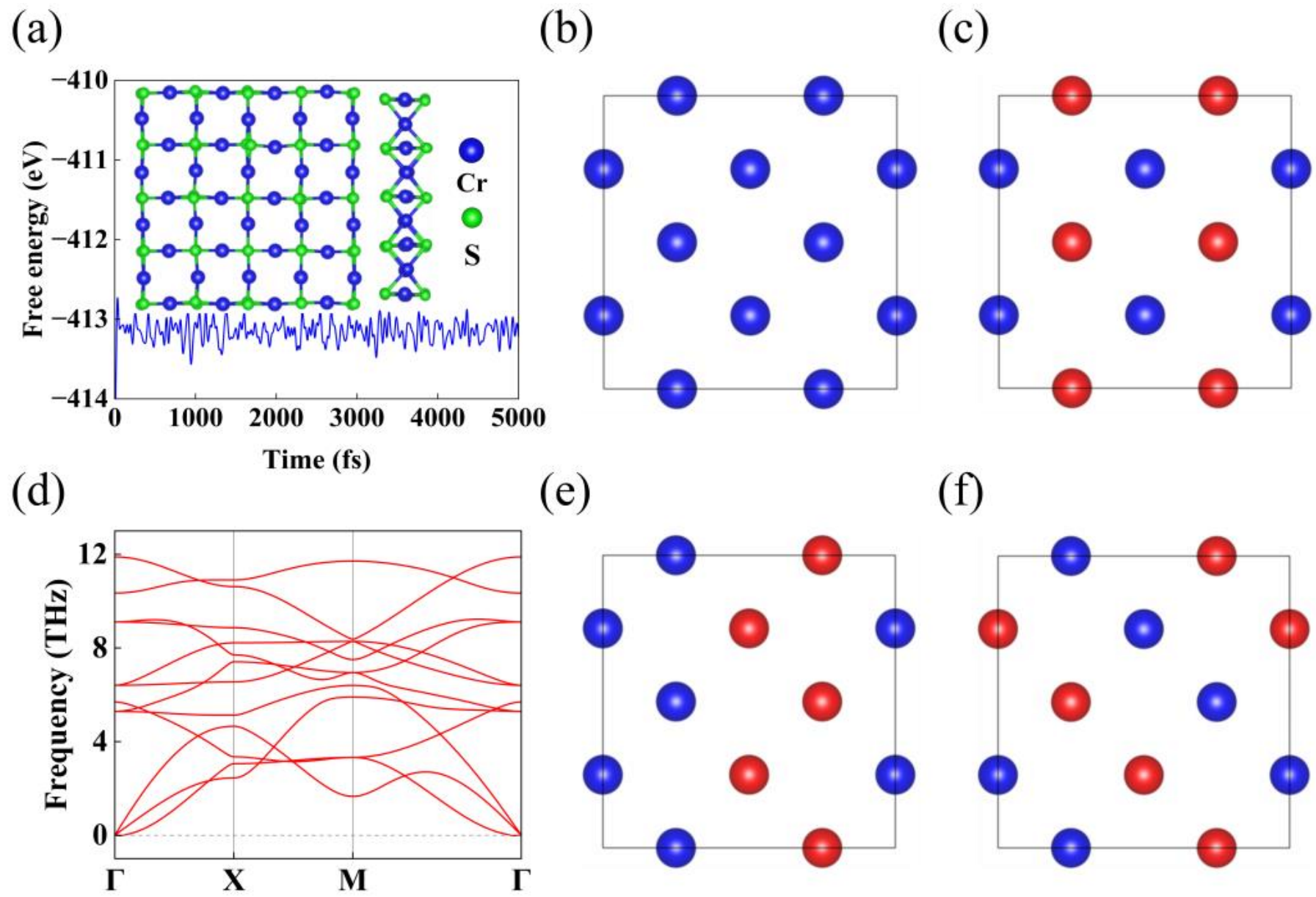

FIG. 2 (a) AIMD simulation of monolayer $Cr_2S_2$ at 300 K, with the inset showing the optimized crystal structure. Blue and green spheres denote Cr and S atoms, respectively. (d) Phonon dispersion of monolayer $Cr_2S_2$, confirming its dynamical stability. (b) FM configuration and (c), (e), and (f) three representative AFM configurations constructed in a 2×2×1 supercell. In the magnetic configurations, blue and red spheres denote Cr atoms with opposite spin orientations.

We then determined the magnetic ground state of monolayer $Cr_2S_2$ by considering one ferromagnetic and three antiferromagnetic configurations in a 2×2×1 supercell, as shown in Figs. 2(b), 2(c), 2(e), and 2(f). The total energy comparison confirms that the $AFM_1$ configuration is the magnetic ground state. Taking the energy of $AFM_1$ as zero, the relative energies of the $AFM_2$, $AFM_3$, and FM states are 215.7, 196.0, and 542.0 meV/Cr atom, respectively. These results demonstrate that monolayer $Cr_2S_2$ is thermally and dynamically stable and possesses a well defined antiferromagnetic ground state, providing a robust material basis for realizing the FQFE-AM coupling and the associated polarization switchable spin-valley physics discussed below.

We next demonstrate the above symmetry mechanism in monolayer $Cr_2S_2$. Figure 3(a) shows the crystal structure of the $h_1$ state. Owing to the combined spin-space and crystal symmetry $[C_2||M_{xy}]$, which protects the valley degeneracy while allowing momentum dependent AM spin splitting, the conduction and valence band extrema at the X and Y valleys remain degenerate but carry opposite spin characters. As shown in the spin-resolved band structure in Fig. 3(d), the X and Y valleys carry opposite spin characters, giving rise to spin-valley locking in the $h_1$ state. Figure 3(b) illustrates the atomic displacement pattern connecting the two FQFE states $h_1$ and $h_2$. The switching process is dominated by fractional lattice displacements of S atoms, as indicated by the red arrows. The corresponding energy profile and Berry-phase polarization along the switching path are shown in Fig. 3(e). A finite energy barrier separates $h_1$ and $h_2$, confirming the bistability of the two FQFE states. Meanwhile, the Berry-phase polarization evolves continuously from one polarization branch to the other, and the two endpoint states exhibit polarizations with equal magnitude but opposite sign, demonstrating switchable FQFE polarization in monolayer $Cr_2S_2$.

According to the modern theory of polarization, the polarization quantum along the [110] direction is given by $Q = e\alpha/\Omega$, where $e$ is the elementary charge, $\alpha$ is the lattice vector along the displacement direction, and Ω is the effective cell volume [36,38]. For monolayer $Cr_2S_2$, the effective thickness is taken as approximately 6.51 Å, yielding a polarization quantum of 91.95 $uC/cm^2$. The Berry-phase polarizations of the $h_1$ and $h_2$ states are approximately +91.95 and −91.95 μC/cm$^2$, respectively, corresponding to a polarization difference of about 2Q. This agreement between the Berry-phase calculation and the polarization-quantum analysis confirms the fractional quantum nature of the $h_1$ to $h_2$ FQFE switching.

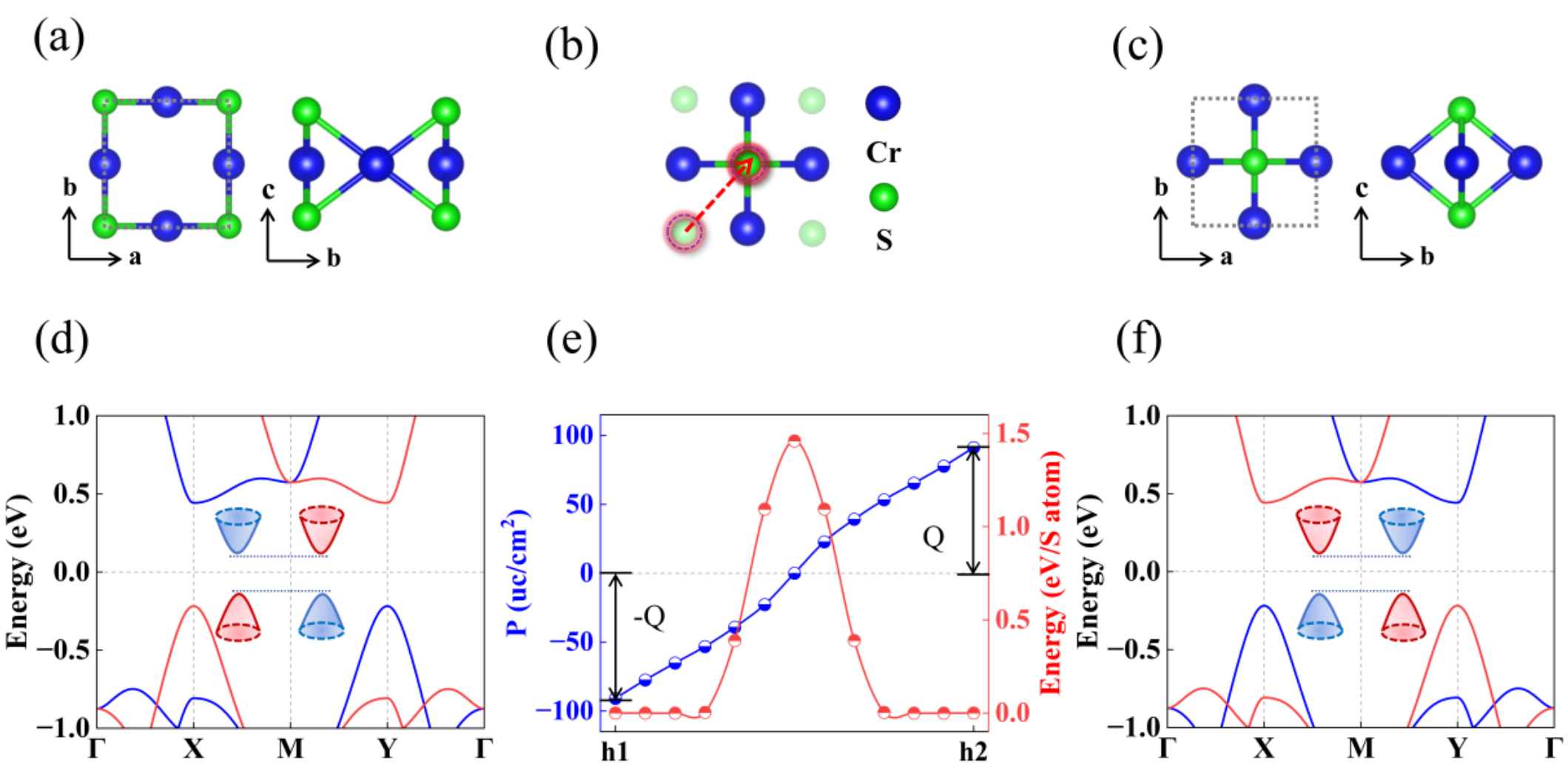

FIG. 3. (a) Top and side views of the $h_1$ state. (b) Atomic displacement pattern for the $h_1$ to $h_2$ switching, with red arrows indicating the fractional lattice displacements of S atoms. (c) Top and side views of the $h_2$ state. (d) Spin-polarized band structure of $h_1$, where blue and red bands denote opposite spin channels. The inset schematics show the spin characters of the X and Y valleys. (e) Energy profile (red curve, right axis) and Berry-phase polarization (blue curve, left axis) along the $h_1$ to $h_2$ switching path. The polarization changes continuously from -Q to +Q. (f) Spin-polarized band structure of $h_2$, showing the interchanged spin characters of the X and Y valleys compared with $h_1$.

Importantly, the fractional atomic displacement switches the polarization without changing the crystallographic symmetry: both $h_1$ and $h_2$ belong to the same space group, as shown in Figs. 3(a) and 3(c). This feature distinguishes FQFE switching from conventional ferroelectric switching driven by symmetry-breaking polar distortions. The band structure of $h_2$ state is shown in Fig. 3(f). Compared with $h_1$ state, the spin-polarized band structure of $h_2$ state is reversed, while the overall valley structure is preserved. In both the conduction and valence bands, the spin characters at the X and Y valleys are interchanged. Since the Néel vector remains unchanged during the switching process, this reversal originates from the FQFE displacement rather than magnetic reorientation. These results demonstrate that monolayer $Cr_2S_2$ realizes a two dimensional FQMF state in which FQFE polarization switching not only controls AM spin splitting but also switches the X/Y spin-valley locking pattern.

We further investigate the Berry-curvature distribution of monolayer $Cr_2S_2$ to reveal the FQFE polarization controlled valley Hall response. The Berry curvature was calculated using the Kubo formula [39],

$$\Omega_z(\mathrm{k}) = -\sum_{\mathrm{n}} \sum_{\mathrm{n}\neq \mathrm{m}} f_{\mathrm{n}}(\mathrm{k}) \frac{2\,\mathrm{Im}\langle \psi_{\mathrm{nk}} | \upsilon_{\mathrm{x}} | \psi_{\mathrm{mk}} \rangle \langle \psi_{\mathrm{mk}} | \upsilon_{\mathrm{y}} | \psi_{\mathrm{nk}} \rangle}{(\mathrm{E}_{\mathrm{nk}} - \mathrm{E}_{\mathrm{mk}})^2} \tag{3}$$

where $f_n(k)$ is the Fermi-Dirac distribution function, $\psi_{nk}$ ($\psi_{mk}$) the Bloch wave function with eigenvalue $E_{nk}$ ($E_{mk}$), $\hat{\upsilon}_{x/y}$ is the velocity operator along the *x*/*y* direction, respectively. As shown in Fig. 4(a), the Berry curvature of the $h_1$ state is mainly concentrated around the X and Y valleys and exhibits opposite signs at the two valleys. Specifically, it is positive at the X valley and negative at the Y valley, with comparable magnitudes. After switching the FQFE state from $h_1$ to $h_2$, the Berry curvature distribution is reversed, as shown in Fig. 4(d): the X valley becomes negative, while the Y valley becomes positive. This FQFE polarization reversible

Berry-curvature distribution provides the microscopic basis for a switchable valley Hall effect. The transport consequence of this Berry-curvature distribution can be understood from the semiclassical anomalous velocity [40], $v \approx -\frac{e}{h} E \times \Omega(\mathrm{k})$, where $E$ is the applied in-plane electric field. Because the X and Y valleys possess opposite Berry curvatures, carriers from these two valleys acquire opposite transverse anomalous velocities under the electric field. In the electron-doped $h_1$ state, electrons occupying the X and Y valleys are deflected toward opposite sample edges. Since these two valleys also carry opposite spin characters due to spin-valley locking, the transverse response is simultaneously valley-contrasting and spin-polarized, as schematically illustrated in Fig. 4(b). A similar Berry curvature driven transverse separation occurs in the hole-doped $h_1$ state, as shown in Fig. 4(c).

Upon switching from $h_1$ to $h_2$ states, both the Berry curvature signs and the valley spin characters are reversed. Consequently, the transverse deflection directions and the associated spin-valley accumulations are switched for both electron and hole doping, as illustrated in Figs. 4(e) and 4(f). Thus, the two nonvolatile FQFE states, combined with electron and hole doping, enable four distinct valley Hall response configurations, each characterized by a different combination of transverse valley deflection and spin-valley locking pattern. These results demonstrate that FQFE polarization switching in monolayer $Cr_2S_2$ provides a symmetry-based route to controlling Berry curvature-driven valley Hall effects in two dimensional FQMFs.

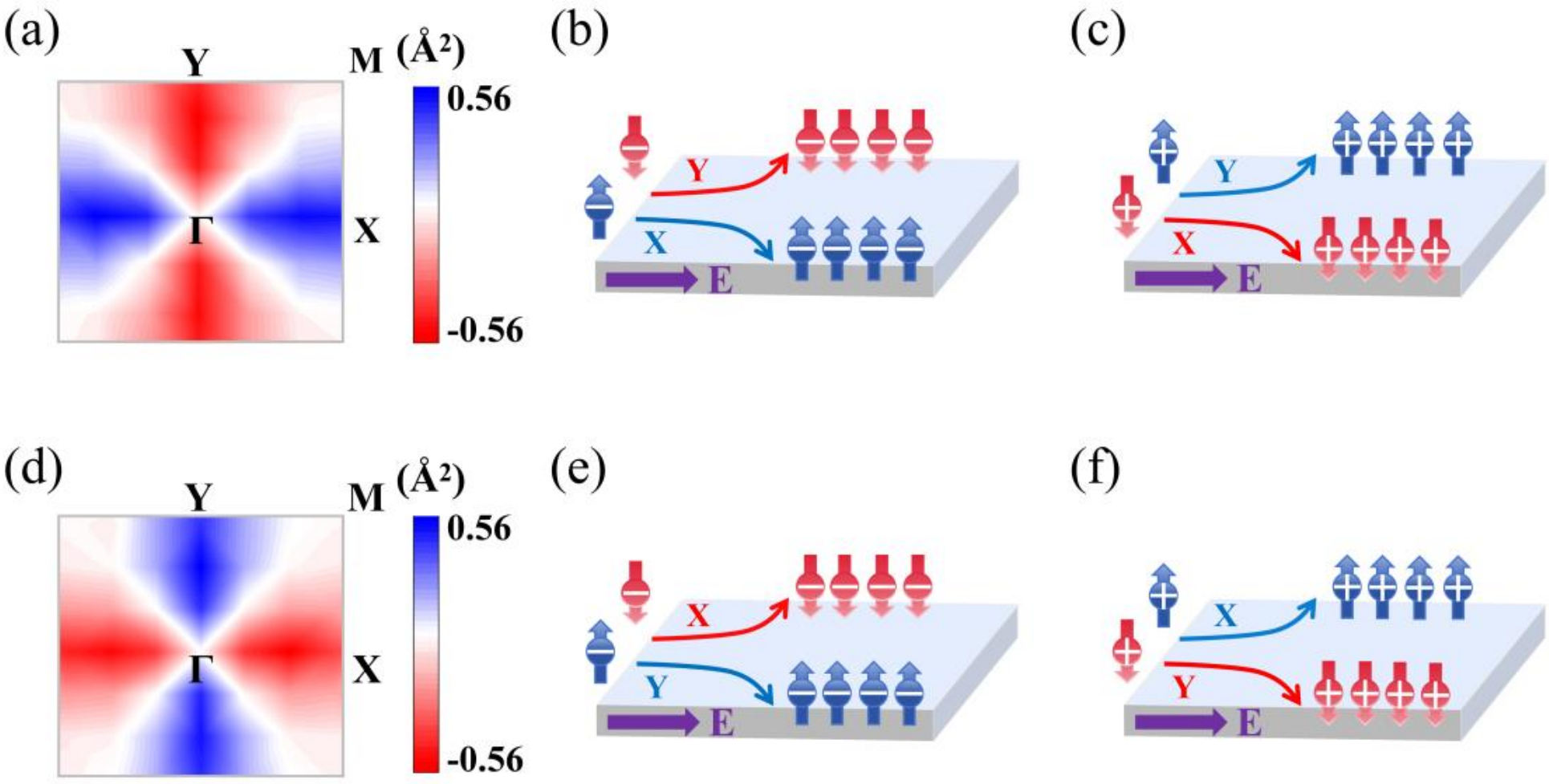

FIG. 4. (a,d) Berry-curvature distribution summed over the occupied valence bands in the $h_1$ and $h_2$ states, respectively. (b,c) Schematic illustrations of the valley Hall effect in the $h_1$ state under electron and hole doping, respectively. (e,f) Schematic illustrations of the valley Hall effect in the $h_2$ state under electron and hole doping, respectively. The Berry curvature at the X and Y valleys reverse their signs upon switching from $h_1$ to $h_2$, leading to polarization-switchable valley-dependent transverse deflection.

*Conclusion.* In conclusion, we demonstrated that monolayer $Cr_2S_2$ serves as a two dimensional FQFE-AM platform in which FQFE and AM are coupled through lattice-symmetry operations. First-principles calculations identify two switchable FQFE states, $h_1$ and $h_2$, which are connected by composite symmetry operations involving a fractional lattice translation combined with $\mathcal{T}$ or $\mathcal{PT}$. This symmetry relation enables FQFE switching to reverse the AM spin-polarized band structure without rotating the Néel vector. As a result, the spin characters of the X and Y crystal valleys are interchanged, realizing nonvolatile polarization switchable spin-valley locking. Furthermore, the two FQFE states exhibit reversed Berry-curvature distributions, leading to polarization controlled valley Hall effects under both electron and hole doping. These findings establish a direct link among FQFE switching, AM spin splitting, spin-valley locking, and Berry curvature driven valley transport, highlighting FQFE-AM coupling as a versatile mechanism for designing electrically programmable spin-valley functionalities in two dimensional materials.

*Acknowledgments.* The work was supported by the National Natural Science Foundation of China (Grants No. 12364007).

*Data availability.* The data that support the findings of this article are openly available from corresponding author upon reasonable request.